\documentclass[aps,pra,preprint,showpacs,amsmath,amssymb]{revtex4}
\usepackage{graphicx}
\usepackage{bm}
\begin{document}

\title{Quantum key distribution using quantum Faraday rotators}

\author{Taeseung Choi} 
\affiliation{Department of Physics, Korea University, Seoul 136-713,
  Korea}
\author{Mahn-Soo Choi}
\email{choims@korea.ac.kr}
\affiliation{Department of Physics, Korea University, Seoul 136-713,
  Korea}
\date{\today}

\begin{abstract}
We propose a new quantum key distribution (QKD) protocol based on the
fully quantum mechanical states of the Faraday rotators.
The protocol is unconditionally secure against collective attacks for
multi-photon source up to two photons on a noisy environment.  It is
also robust against impersonation attacks.
The protocol may be implemented experimentally with the current
spintronics technology on semiconductors.
\end{abstract}
\maketitle

\newcommand\ket[1]{\left|{\textstyle#1}\right\rangle}
\newcommand\bra[1]{\left\langle{\textstyle#1}\right|}
\newcommand\braket[1]{\left\langle{\textstyle#1}\right\rangle}

\section{Introduction}
\label{sec:introduction}

The computational algorithm powered by quantum mechanics, on the one
hand, has posed a serious threat to the classical
cryptosystem\cite{Shor94a}.
On the other hand, quantum cryptography allows for secure sharing of
private keys.  Ever since the pioneering works by Bennet, Brassard, and
Ekert\cite{Bennett84a,Ekert91a,Bennett92c}, a great number of new
quantum key distribution (QKD) protocols have been proposed to enhance
the security and efficiency under non-idealistic situations and
to incorporate new ideas\cite{Gisin02a}.
In particular, Bostr\"om and Felbinger\cite{Bostrom02a} recently
proposed the so-called \emph{ping-pong protocol}.  The protocol is
interesting in that it enables direct communication deterministically
and without classical communications (except for checking
eavesdropping).  Although the original protocol turned out to be
insecure in the case of lossy channels\cite{Wojcik03a} and against blind
attacks without eavesdropping\cite{Cai03a}, the idea still survives in a
recent modified version\cite{Lucamarini05a}.

In the ping-pong protocol\cite{Bostrom02a,Lucamarini05a}, Bob sends a
qubit to Alice, Alice performs a unitary operation on it with a random
probability $p$ and send it back to Bob, and finally Bob make a
measurement on it.  The unitary operation by Alice (if ever performed)
transforms the initial state of the qubit to a state orthogonal to the
initial state.  This enables Bob to read Alice's massage directly.
Putting another way, the unitary operation is performed conditioned on
the \emph{classical} information ($0$ or $1$) that Alice wants to send
to Bob.
A conceptually interesting question would be, ``What if we perform the
unitary operation conditioned on the \emph{quantum state} of another
qubit?''  In this work, we propose a new QKD protocol implementing this
idea and address the security issues of it.  The protocol is explained
in Section~\ref{sec:protocol}.  We will show in
Sections~\ref{sec:securityProof} and \ref{sec:impersonation} that the
protocol is secure against eavesdropping for ideal single-photon source
and robust against impersonation attacks.  The protocol turns out to be
insecure when the photon source produces more
than two photons; this will be analyzed in Section~\ref{sec:PNS}.  We
will discuss in Section~\ref{sec:experiment} possible experimental
realizations of the protocol using semiconductor spintronics.

\begin{figure}
\centering
\includegraphics*[width=3cm]{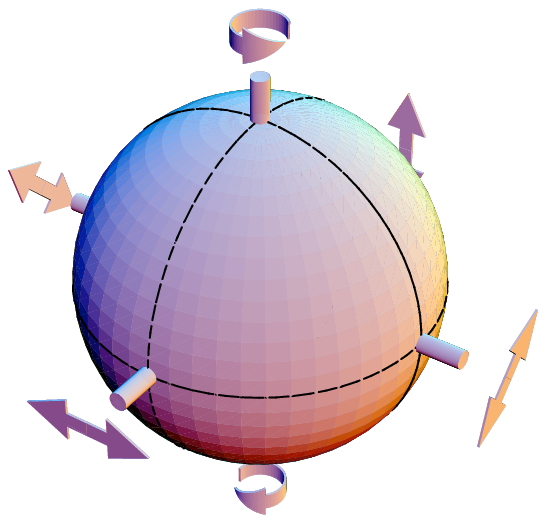}
\includegraphics*[width=5cm]{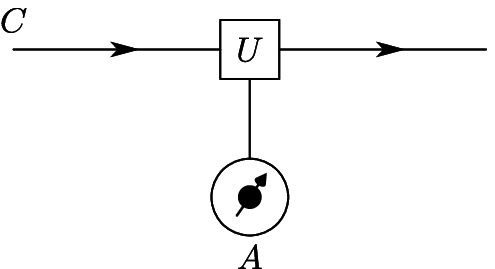}
\caption{(color on-line) (a) Poincar\'e sphere for photon polarization or
  Bloch sphere for spin.  (b) Quantum Faraday rotation (QFR) or
  conditional rotation $U_{A;C}$ on $C$ conditioned by $A$; see
  Eq.~(\ref{paper::eq:3}).  It rotates the state of qubit $C$ around
  $z$-axis by angle $\pm\pi/2$ depending on the state of qubit $A$.}
\label{fig:1}
\end{figure}

\begin{figure}
\centering
\includegraphics*{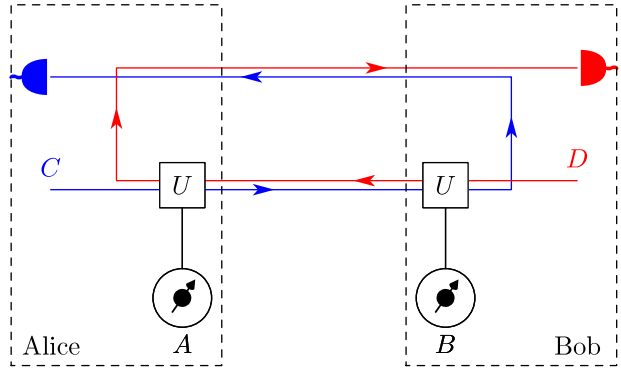}
\caption{(color on-line) Quantum key distribution protocol using quantum
  Faraday rotators.}
\label{fig:2}
\end{figure}

\section{Protocol}
\label{sec:protocol}
While the protocol is independent of the physical system in use, we will
have in mind the photon polarizations as travel qubits and electron
spins as home qubits.  In the description of the protocol, we will use
as the basis the eigenstates of $\sigma^z$, $\ket{\uparrow}$ (right-handed
circular polarization) and
$\ket{\downarrow}$ (left-handed circular polarization).
We denote by $\ket{\phi}$ the state along the azimuthal angle $\phi$ on the
equator of the Poincar\'e (or Bloch) sphere:
\begin{equation}
\label{paper::eq:2}
\ket\phi = \frac{\ket\uparrow + e^{+i\phi}\ket\downarrow}{\sqrt{2}} \,.
\end{equation}
The key element of our protocol will be the \emph{quantum Faraday
  rotation} (QFR), namely, the Faraday rotation by angle $\pi/2$ around
$z$-axis of the Poincar\'e sphere
\begin{equation}
\label{paper::eq:3}
U_{A;C}=\exp\left[-i(\pi/4)\sigma_A^z\sigma_C^z\right]
\end{equation}
on the travel qubit $C$ \emph{conditioned by} the home qubit $A$.
For example, operating on the product state $\ket{\phi=0}_A\ket{\phi}_C$,
it gives
\begin{equation}
\label{paper::eq:1}
U_{A;C}\ket{0}_A\ket{\phi}_C =
\frac{e^{-i\pi/4}\ket\uparrow_A\ket{\phi_+}_C
  + e^{+i\pi/4}\ket\downarrow_A\ket{\phi_-}_C}{\sqrt{2}},
\end{equation}
where $\ket{\phi_\pm}=\ket{\phi\pm\pi/2}$.
In the quantum information theoretic terms, the QFR in
Eq.~(\ref{paper::eq:3}) corresponds to the \emph{conditional phase
  shift}.  Possible physical realizations of QFR will be discussed
later.

The protocol is as following.  (1) To start the $n$th iteration of the
protocol, Alice and Bob first prepare their home qubits $A$ and $B$,
respectively, in the state $\ket{\phi=0}$\cite{endnote:1}.  (2) Alice
then takes a travel qubit $C$ and prepares it in the state $\ket\alpha$.
The angle $\alpha$ should be chosen randomly in the interval
$0\leq\alpha<2\pi$.  (3) Alice performs (by interacting $A$ and $C$) the
QFR $U_{A;C}$ on $C$ and send it to Bob.  We note that on its way to
Bob, the travel qubit $C$ is maximally entangled with $A$:
\begin{equation}
\label{paper::eq:5}
e^{-i\pi/4}\ket\uparrow_A\ket{\alpha_+}_C
  + e^{+i\pi/4}\ket\downarrow_A\ket{\alpha_-}_C
\end{equation}
(not normalized).  (4) Bob receives $C$, performs $U_{B;C}$ on it, and
send it back to Alice.  The qubit $C$ is again maximally entangled on
its way back to Alice, now with both $A$ and $B$ :
\begin{equation}
\label{paper::eq:4}
\left(\ket{\uparrow\downarrow}+\ket{\downarrow\uparrow}\right)_{AB}
\ket{\alpha}_C
-i\left(\ket{\uparrow\uparrow}-\ket{\downarrow\downarrow}\right)_{AB}
\ket{\bar\alpha}_C
\end{equation}
(not normalized), where $\ket{\bar\alpha}\equiv\ket{\alpha+\pi}$.  (5)
Now Bob takes his own travel qubit $D$ and prepares it in the state
$\ket\beta$.  The angle $\beta$ should be chosen randomly in the
interval $0\leq\beta<2\pi$.  (6) Bob performs the QFR $U_{B;D}$ on $D$
and send it to Alice.  (7) Alice receives $D$, performs $U_{A;D}$ on it,
and send it back to Bob.  The final state of all the qubits $A$, $B$,
$C$, and $D$ is given by a GHZ-like state
\begin{equation}
\label{paper::eq:6}
\left(\ket{\uparrow\downarrow}+\ket{\downarrow\uparrow}\right)_{AB}\ket{\alpha\beta}_{CD}
-\left(\ket{\uparrow\uparrow}+\ket{\downarrow\downarrow}\right)_{AB}\ket{\bar\alpha\bar\beta}_{CD}
\end{equation}
(8) Alice measures the observable
$S_\alpha=\cos\alpha\sigma^x+\sin\alpha\sigma^y$ on $C$.  Likewise, Bob
measures the observable $S_\beta=\cos\beta\sigma^x+\sin\beta\sigma^y$ on
$D$.  They will get (in the ideal case) the identical result $+1$ or
$-1$, which enables Alice and Bob to share the the key $K_{2n-1}=1$ or
$0$.  (9) If $K_{2n-1}=1$, Bob performs $\sigma^x$ (the NOT gate), on
his home qubit $B$.  (10) Alice and Bob measures $\sigma^z$ on their
home qubits $A$ and $B$, respectively.  Depending on the measurement
result, another bit of key $K_{2n}=0$ ($\sigma^x=+1$) or $1$
($\sigma^x=-1$) is generated. (11) Repeat the steps 1 through 10 with
$n$ increased by $1$ until $n$ becomes $N$. (12) Alice and Bob takes
randomly $M$ bits out of $\{K_{2k-1}|k=1,\cdots,N\}$, and test possible
eavesdropping (or any other attack) by comparing the values through a
classical communication channel.

A few remarks on the procedure are in order.
Alice can measure $S_\alpha$ (see Step 8 above) even before the Step 5.
It follows from the GHZ-like structure of the states in
Eqs.~(\ref{paper::eq:4}) and (\ref{paper::eq:6}).  Step 9 is not
essential.  It can be removed with a minor change in Step 10.

Before analyzing the security of the protocol, we point out a few
interesting features of the protocol.  First, the travel qubit is always
in a maximally entangled states with the home qubit(s) whenever exposed
to eavesdropping.  This is the essential feature of the protocol that
provides the protocol with the security.  Second, at the key sharing
stage no classical communication is necessary.  The key is shared only
through the quantum channel\cite{endnote:2}.  This is also closely
related to the security of the protocol.  Third, two bits are generated
out of one iteration and they have the common security fate.  If the
first bit has been tampered by eavesdropping or noise in the channel,
the security of the second bit is not guaranteed either.

\begin{figure}
\centering
\includegraphics*{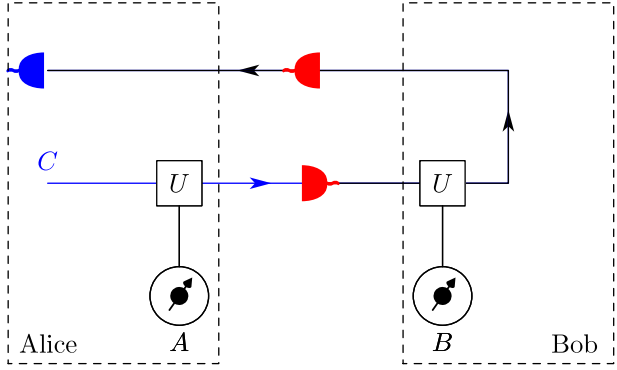}
\caption{(color on-line) General attack on a noisy environment.}
\label{fig:3}
\end{figure}

\section{Security Proof}
\label{sec:securityProof}
Let us analyze general attacks from a third party in case the photon
source generates single photon. We closely follow the lines in
Ref.\cite{Lucamarini05a}. As usual, Eve is assumed to be an almighty
eavesdropper limited only by the law of physics.  The most general
(assuming that Eve does not know Alice's choice of basis) operation
$\hat{\cal E}_1$ Eve can do on the travel qubit $C$ can be written as
\begin{equation}
\label{Eq:generalAttack}
\hat{\cal E}_1 \ket{\gamma}_C \ket{\epsilon}_E
= e \ket{\gamma}_C \ket{\epsilon_{00}}_E+ 
f \ket{\bar\gamma}_C \ket{\epsilon_{01}}_E,
\end{equation}
and
\begin{equation}
\label{qkd::eq:1}
\hat{\cal E}_1 \ket{\bar\gamma}_C \ket{\epsilon}_E = 
e \ket{\bar\gamma}_C \ket{\epsilon_{11}}_E + f \ket{\gamma}_C
\ket{\epsilon_{10}}_E,
\end{equation}
where the states $\ket{\epsilon_{00}}$, $\ket{\epsilon_{01}}$,
$\ket{\epsilon_{11}}$, and $\ket{\epsilon_{10}}$ of the ancilla $E$ are
normalized, but not orthogonal to each other.
Without loss of generality, we can set
\begin{math}
\braket{\epsilon_{00}|\epsilon_{01}} =
\braket{\epsilon_{00}|\epsilon_{10}} =
\braket{\epsilon_{10}|\epsilon_{11}} =
\braket{\epsilon_{01}|\epsilon_{11}} = 0
\end{math},
from the unitarity of $\mathcal{E}_1$\cite{Lucamarini05a}.

The basis $\{\ket{\gamma},\ket{\bar\gamma}\}$ for $C$ is an arbitrary
choice made by Eve. Recall that the angle $\alpha$ has been chosen
randomly for each travel qubit $C$ and is never announced to the public;
this is one of the biggest differences of our protocol both from the
BB84-type and ping-pong-type protocols.

When
\begin{math}
\braket{\epsilon_{00}|\epsilon_{11}} =
\braket{\epsilon_{01}|\epsilon_{10}} = 1
\end{math}, Eve cannot distinguish between $\ket{\gamma}_C$ and
$\ket{\bar\gamma}_C$ by any measurement on her ancillae.  In this case,
Eve can acquire no more information than no attack is performed.
Therefore, a minimal requirement for Eve's strategy is that such an
operation as gives no information at all to her should not be detected
by the legitimate partners (Alice and Bob).
This can be achieved if Eve does not disturb travel qubits.  It gives
the condition, $|e|=1$ and $|f|=0$.
In passing, we note that $\braket{\epsilon_{00}|\epsilon_{11}}=0$
corresponds to an intercept-and-resend attack.

Having this ($|e|=1$ and $|f|=0$) in mind, we rewrite the attack operation
on the travel qubit as
\begin{multline}
\label{Eq:AttackTravel}
\hat{\cal E}_1 \ket{\alpha_+}_C
= \ket{\alpha_+}_C 
\left(\cos^2{\frac{\tilde{\alpha}}{2}} \ket{\epsilon_{00}}
  + \sin^2{\frac{\tilde{\alpha}}{2}} \ket{\epsilon_{11}}\right) \\\mbox{}
+ i \sin{\frac{\tilde{\alpha}}{2}}
\cos{\frac{\tilde{\alpha}}{2}}\ket{\alpha_-}_C
\left(\ket{\epsilon_{00}} -  \ket{\epsilon_{11}}\right)
\end{multline}
and
\begin{multline}
\hat{\cal E}_1 \ket{\alpha_-}_C
= -i \sin{\frac{\tilde{\alpha}}{2}} \cos{\frac{\tilde{\alpha}}{2}} 
\ket{\alpha_+}_C \left(\ket{\epsilon_{00}}
  -  \ket{\epsilon_{11}}\right) \\\mbox{}
+ \ket{\alpha_-}_C
\left(\sin^2{\frac{\tilde{\alpha}}{2}} \ket{\epsilon_{00}}
  + \cos^2{\frac{\tilde{\alpha}}{2}} \ket{\epsilon_{11}}\right),
\end{multline}
where $\tilde\alpha\equiv\alpha-\gamma+\pi/2$.

On $C$'s way from Bob back to Alice, Eve can perform another similar
attack $\hat{\cal E}_2$ with a new ancilla $F$.  With
the same requirement as in $\hat{\cal E}_1$, the attack operation
$\hat{\cal E}_2$ takes the simple form
\begin{eqnarray}
\label{Eq:AttackReturn}
\hat{\cal E}_2 \ket{\gamma}_C \ket{\eta}_F =  \ket{\gamma}_C
\ket{\eta_{00}}_F
\end{eqnarray}
and
\begin{equation}
\hat{\cal E}_2 \ket{\bar\gamma}_C \ket{\eta}_F = 
 \ket{\bar\gamma}_C \ket{\eta_{11}}_F .
\end{equation}

Since our protocol is symmetric between Alice and Bob, Eve's attack
operations $\hat{\cal E}_1'$ and $\hat{\cal E}_2'$ on Bob's travel qubit $D$
can be written, analogously to $\hat{\cal E}_1$ and $\hat{\cal E}_2$,
with respect to new ancillae $E'$ and $F'$.
The optimal Eve's attack will be the symmetric one such that
\begin{math}
\braket{\epsilon_{00}|\epsilon_{11}}=
\braket{\epsilon_{00}'|\epsilon_{11}'}
\end{math}
and
\begin{math}
\braket{\eta_{00}|\eta_{11}}=
\braket{\eta_{00}'|\eta_{11}'}
\end{math}.
The angle $\tilde\beta=\beta-\gamma+\pi/2$ relates
Bob's choice $\{\ket{\beta},\ket{\bar\beta}\}$ and Eve's choice
$\{\ket{\gamma},\ket{\bar\gamma}\}$ for the basis for $D$.

After all the procedures by Alice and Bob, Eve performs a collective
measurement on her ancillae $E$, $F$, $E'$, and $F'$.  From the
measurement result, she extracts the information about the state of
Alice's home qubit $A$ and Bob's $B$; namely, the information about the
results of the QFR on the travel qubits $C$ and $D$.  The information is
eventually the information about the key values shared by Alice and Bob.

The operations $\hat{\cal E}_1$, $\hat{\cal E}_2$, $\hat{\cal E}_1'$, and
$\hat{\cal E}_2'$ by Eve inevitably disturb the quantum state of the travel
qubit $C$ and $D$.  Simply comparing the test key bits (step 12 of the
protocol), Alice and Bob may detect the attack.  The detection
probability $p_d$ depends on the angle differences $\tilde{\alpha}$ and
$\tilde{\beta}$.
Since the angles $\tilde{\alpha}$ and $\tilde{\beta}$ are randomly
distributed, the detection probability is given by
\begin{equation}
\label{Eq:AverageDetect}
p_d = \frac{3}{8}- \frac{1}{8}
\left(\cos^2{x} + \cos^2{y} +\cos^2{x}\cos^2{y}\right) \,,
\end{equation}
where
\begin{math}
\cos{x}\equiv\braket{\epsilon_{00}|\epsilon_{11}}
=\braket{\epsilon_{00}'|\epsilon_{11}'}
\end{math}
and
\begin{math}
\cos{y}\equiv\braket{\eta_{00}|\eta_{11}}
=\braket{\eta_{00}'|\eta_{11}'}
\end{math}.
The maximum value of $p_d$ is 3/8 corresponding to the
intercept-and-resend attack($\cos{x}=\cos{y}=0$).

Let us suppose that the initial state prepared by Alice and Bob is given by
\begin{eqnarray}
\label{Eq:InitialAB}
\ket{\Psi}_{i} = \ket{0}_A \ket{0}_B \ket{\alpha \beta}_{CD}.
\end{eqnarray} 
After all attacks the final state is given by 
\begin{widetext}
\begin{multline}
\label{Eq:AttackFinal}
\frac{1}{{2}} \ket{\alpha\beta}_{CD} \Bigg\{
\frac{1}{4}\sin\tilde\alpha\sin\tilde\beta
\ket{\uparrow\uparrow}_{AB}\ket{1}_{EF}\ket{1'}_{E'F'}
+ \ket{\uparrow \downarrow}_{AB}\ket{5}_{EF}\ket{2'}_{E'F'} \\\mbox{}
+ \ket{\downarrow \uparrow}_{AB}\ket{2}_{EF}\ket{5'}_{E'F'}
+ \frac{1}{4}\sin\tilde\alpha\sin\tilde\beta
\ket{\downarrow\downarrow}_{AB}\ket{4}_{EF}\ket{4'}_{E'F'}\Bigg\} \\\mbox{}
+ \frac{1}{{2}}\ket{\alpha\bar\beta}_{CD} \Bigg\{
- \frac{i}{2}\sin\tilde\alpha
\ket{\uparrow\uparrow}_{AB}\ket{1}_{EF}\ket{3'}_{E'F'}
- \frac{i}{2}\sin\tilde\beta
\ket{\uparrow\downarrow}_{AB}\ket{5}_{EF}\ket{1'}_{E'F'} \\\mbox{}
+ \frac{i}{2}\sin\tilde\beta
\ket{\downarrow\uparrow}_{AB}\ket{2}_{EF}\ket{4'}_{E'F'}
+ \frac{i}{2}\sin\tilde\alpha
\ket{\downarrow\downarrow}_{AB}\ket{4}_{EF}\ket{6'}_{E'F'} \Bigg\} \\\mbox{}
\frac{1}{{2}}\ket{\bar\alpha\beta}_{CD} \Bigg\{
- \frac{i}{2}\sin\tilde\beta
\ket{\uparrow\uparrow}_{AB}\ket{3}_{EF}\ket{1'}_{E'F'}
+ \frac{i}{2}\sin\tilde\alpha
\ket{\uparrow\downarrow}_{AB}\ket{4}_{EF}\ket{2'}_{E'F'} \\\mbox{}
- \frac{i}{2}\sin\tilde\alpha
\ket{\downarrow\uparrow}_{AB}\ket{1}_{EF}\ket{5'}_{E'F'}
+ \frac{i}{2}\sin\tilde\beta
\ket{\downarrow\downarrow}_{AB}\ket{6}_{EF}\ket{4'}_{E'F'} \Bigg\} \\\mbox{}
+ \frac{1}{{2}}\ket{\bar\alpha\bar\beta}_{CD} \Bigg\{
- \ket{\uparrow\uparrow}_{AB}\ket{3}_{EF}\ket{3'}_{E'F'}
+\frac{1}{4}\sin\tilde\alpha\sin\tilde\beta
\ket{\uparrow\downarrow}_{AB}\ket{4}_{EF}\ket{1'}_{E'F'} \\\mbox{}
+ \frac{1}{4}\sin\tilde\alpha\sin\tilde\beta
\ket{\downarrow\uparrow}_{AB}\ket{1}_{EF}\ket{4'}_{E'F'}
-\ket{\downarrow\downarrow}_{AB}\ket{6}_{EF}\ket{6'}_{E'F'} \Bigg\}
\end{multline}
\end{widetext}
with
\begin{equation}
\ket{1}_{EF} \equiv
\ket{\epsilon_{00}}_E\ket{\eta_{00}}_F
- \ket{\epsilon_{11}}_E\ket{\eta_{11}}_F
\end{equation}
\begin{equation}
\ket{2}_{EF} \equiv
\sin^2{\frac{\tilde{\alpha}}{2}} \ket{\epsilon_{00}}_E\ket{\eta_{00}}_F
+ \cos^2{\frac{\tilde{\alpha}}{2}} \ket{\epsilon_{11}}_E\ket{\eta_{11}}_F
\end{equation}
\begin{equation}
\ket{3}_{EF} \equiv
\cos^2{\frac{\tilde{\alpha}}{2}}\ket{\epsilon_{00}}_E\ket{\eta_{00}}_F
+ \sin^2{\frac{\tilde{\alpha}}{2}}\ket{\epsilon_{11}}_E \ket{\eta_{11}}_F
\end{equation}
\begin{equation}
\ket{4}_{EF} \equiv
\ket{\epsilon_{00}}_E\ket{\eta_{11}}_F
- \ket{\epsilon_{11}}_E\ket{\eta_{00}}_F
\end{equation}
\begin{equation}
\ket{5}_{EF} \equiv
\cos^2{\frac{\tilde{\alpha}}{2}} \ket{\epsilon_{00}}_E\ket{\eta_{11}}_F
+ \sin^2{\frac{\tilde{\alpha}}{2}} \ket{\epsilon_{11}}_E\ket{\eta_{00}}_F
\end{equation}
and
\begin{equation}
\ket{6}_{EF} \equiv
\sin^2{\frac{\tilde{\alpha}}{2}}\ket{\epsilon_{00}}_E\ket{\eta_{11}}_F
+ \cos^2{\frac{\tilde{\alpha}}{2}}\ket{\epsilon_{11}}_E\ket{\eta_{00}}_F
\end{equation}
The states $\ket{1'}_{E'F'}$, $\ket{2'}_{E'F'}$, $\ket{3'}_{E'F'}$,
$\ket{4'}_{E'F'}$, $\ket{5'}_{E'F'}$, and $\ket{6'}_{E'F'}$ are defined
analogously (with $\tilde\alpha$ replaced by $\tilde\beta$).

Equation~(\ref{Eq:AttackFinal}) clearly reveals how Eve can extract the
information about the quantum state of the Alice's and Bob's home qubits
$A$ and $B$, respectively.  For example, Eve can infer the state
$\ket{\uparrow}_B$ on Bob's home qubit $B$ if she finds her ancilla qubits
$E$ and $F$ in the collective state $\ket{1}_{EF}$, $\ket{2}_{EF}$, or
$\ket{3}_{EF}$.  Likewise, Eve infers the state $\ket{\downarrow}_B$ if she
finds $E$ and $F$ in the state $\ket{4}_{EF}$, $\ket{5}_{EF}$, or
$\ket{6}_{EF}$.  The state of Alice's home qubit $A$ can be inferred
analogously from the ancillae $E'$ and $F'$.  The remaining question for
Eve would be, for example, how to distinguish the states $\ket{1}_{EF}$,
$\ket{2}_{EF}$, and $\ket{3}_{EF}$ from $\ket{4}_{EF}$, $\ket{5}_{EF}$, 
and $\ket{6}_{EF}$.
  
To this end, we first note that
\begin{equation}
\braket{1|4}_{EF}=\braket{1|5}_{EF}=\braket{1|6}_{EF}=0
\end{equation}
and that
\begin{equation}
\braket{4|1}_{EF}=\braket{4|2}_{EF}=\braket{4|3}_{EF}=0 \,.
\end{equation}
Therefore, Eve's best policy will be first to exploit the orthogonal
subspaces containing $\ket{1}_{EF}$ and $\ket{4}_{EF}$, respectively,
and then to distinguish the non-orthogonal states, namely $\ket{2}_{EF}$
and $\ket{3}_{EF}$ from $\ket{5}_{EF}$ and $\ket{6}_{EF}$, within these
subspaces.
Further, defining the normalized overlap
\begin{equation}
\overline{\braket{i|j}}_{EF} \equiv
\frac{\braket{i|j}_{EF}}{\sqrt{\braket{i|i}_{EF}\braket{j|j}_{EF}}}
\end{equation}
($i,j=1,\cdots,6$),
we have the inequalities
\begin{equation}
\overline{\braket{2|5}}_{EF} = \overline{\braket{3|6}}_{EF}
\geq \min\{\cos{x},\cos{y}\}
\end{equation}
and
\begin{equation}
\overline{\braket{2|6}}_{EF} = \overline{\braket{3|5}}_{EF}
\geq \min\{\cos{x},\cos{y}\}
\end{equation}
Namely, the states $\ket{2}_{EF}$ and $\ket{3}_{EF}$ can be
distinguished worse from $\ket{5}_{EF}$ and $\ket{6}_{EF}$ than any two
states with the mutual overlap of
\begin{math}
\min\{\cos{x},\cos{y}\}
\end{math}
can be distinguished from each other.  Based on this observation, we
analyze the worst case, where
\begin{math}
\overline{\braket{2|5}}_{EF} = \overline{\braket{3|6}}_{EF} =
\overline{\braket{2|6}}_{EF} = \overline{\braket{3|5}}_{EF} =
\min\{\cos{x},\cos{y}\}.
\end{math}
Further, it is clear that the optimal attack for Eve is the balanced one
\cite{Lucamarini05a}, for which $\cos{x}=\cos{y}$, and hereafter we
focus on the balanced case.

\begin{figure}
\centering
\includegraphics*[width=6cm]{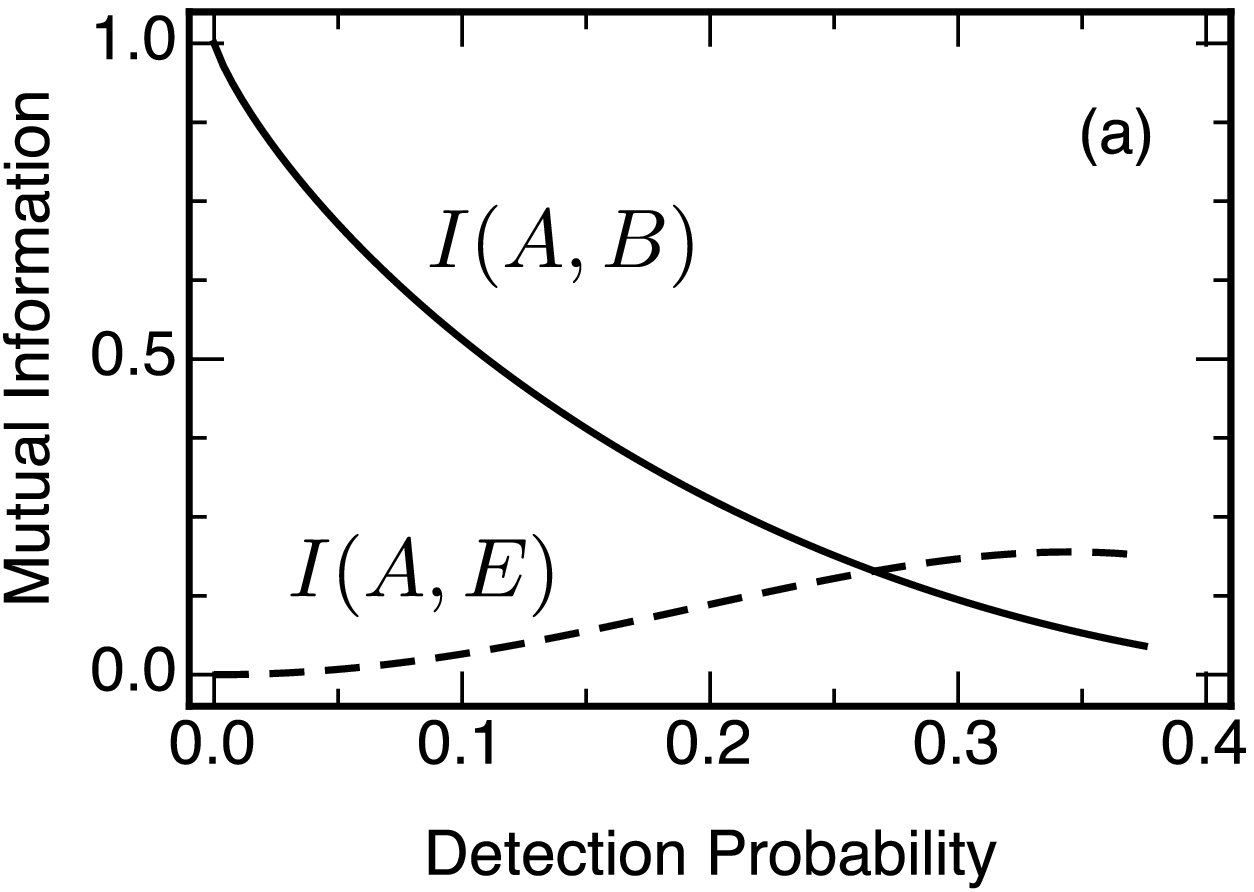}
\includegraphics*[width=6cm]{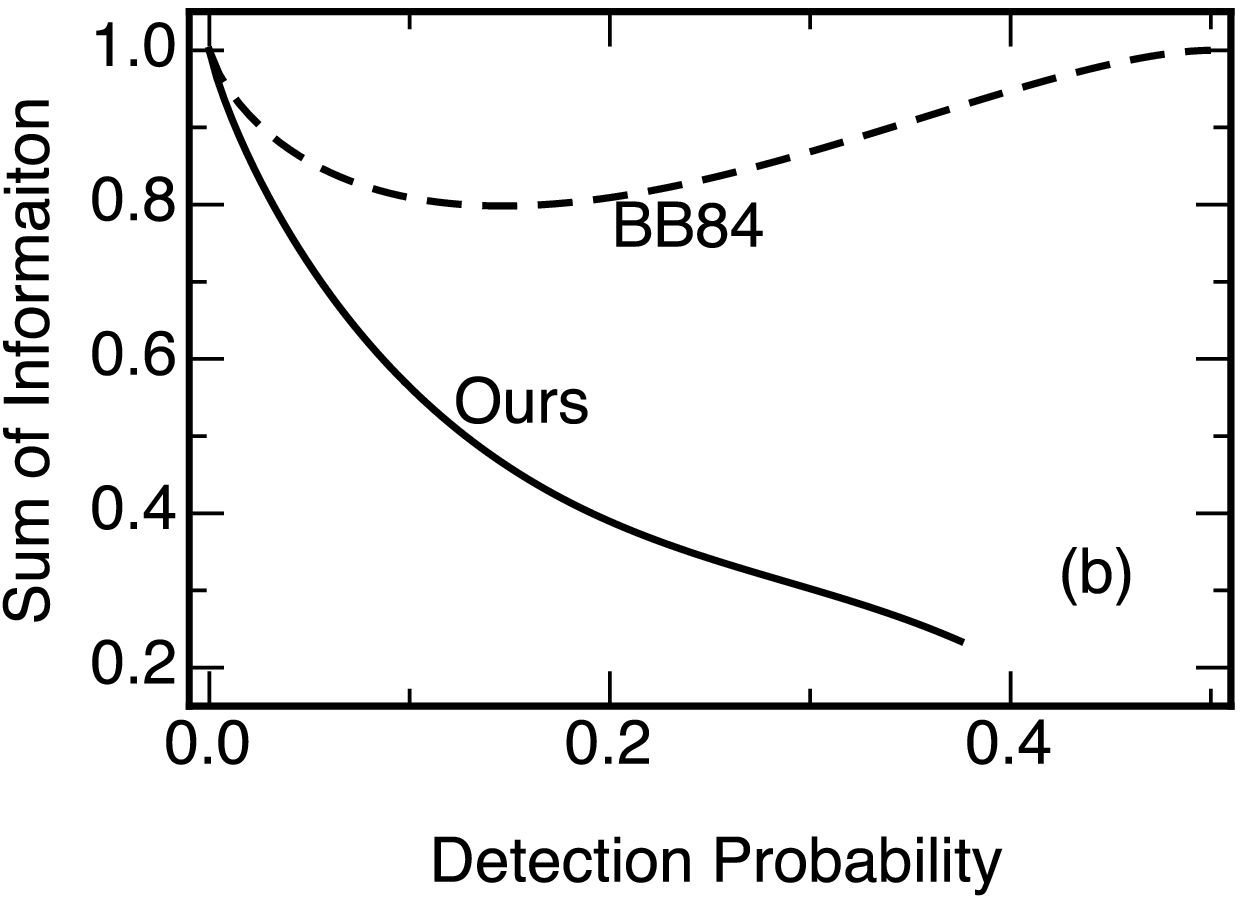}
\caption{(a) Mutual information a function of the detection probability,
  $p_d$, for general incoherent attacks against our protocol. The
  solid (dashed) line represents the mutual information $I(A,B)$
  [$I(A,E)$] between Alice and Bob (Alice and Eve). (b) $I(A,B)+I(A,E)$
  as a function of $p_d$ for our protocol (solid line) and for the
  BB84 protocol (dashed line).  For our protocol, the maximum value of
  $p_d$ is 3/8.}
\label{fig:4}
\end{figure}

Putting all the above observations together and with lengthy algebra,
one can calculate the mutual information $I(A,B)$ between Alice and Bob
and $I(A,E)$ [or $I(B,E)$] between Alice (or Bob) and Eve; note that
because of the symmetry in our protocol, $I(A,E)=I(B,E)$.
They are given by
\begin{equation}
\label{qkd::eq:2}
I(A,B) = 1 + p_d\log_2p_d + (1-p_d)\log_2(1-p_d)
\end{equation}
and
\begin{equation}
\label{qkd::eq:3}
I(A,E) = 1 + p_e\log_2p_e + (1-p_e)\log_2(1-p_e) \,,
\end{equation}
respectively. $p_d$ in Eq.~(\ref{qkd::eq:2}) is the detection
probability [see Eq.~(\ref{Eq:AverageDetect})] for the balanced attack
($\cos{x}=\cos{y}$), and $p_e$ in Eq.~(\ref{qkd::eq:3}) is defined by
\begin{widetext}
\begin{equation}
p_e = \frac{1}{2} - \frac{1}{2}\sqrt{1-2p_d}(1-\sqrt{1-2p_d})
\left[2\sqrt{1-2p_d}+\sqrt{2(1-\sqrt{1-2p_d})}\right].
\end{equation}
\end{widetext}
%
For a QKD to be secure, it is required that $I(A,B)\geq
I(A,E)$\cite{Gisin02a}.
%
The mutual information $I(A,B)$ and $I(A,E)$ are
plotted as functions of the detection probability $p_d$ in
Fig.~\ref{fig:4}.
The maximum information between Alice and Eve occurs at
$p_d=0.345$, which is less than the maximum detection probability
($p_d=3/8$) corresponding to the intercept-and-resend attack.
This means that the intercept-and-resend attack is not an optimal
attack for Eve.
$I(A,B)$ and $I(A,E)$ becomes equal for the detection probability $p_d =
0.266188$.  This detection probability is greater than $p_d=0.18$ for
the ping-pong protocol \cite{Lucamarini05a} and $p_d=0.15$ for BB84
protocol.

So far the security has been analyzed for incoherent attacks.  In
general, Eve can attacks many qubits coherently by collecting many
ancillae and performing a global measurement on them.  Since our
protocol shares many common features with the BB84 or similar protocols,
we can first follow the lines in Section VI.G of Ref.~\cite{Gisin02a} to
prove the security of our protocol against collective
attacks\cite{Biham97a}.  An argument for the security against the most
general coherent attacks\cite{Mayers93a} is given below.  After Alice
and Bob repeats the protocol $n$ times to share a key of length of $2n$
bits, the sum of the mutual information $I(A,B)$ and $I(A,E)$ should be
less than $2n$, i.e.,
\begin{equation}
\label{qkd::eq:4}
I(A,B) + I(A,E) \leq 2n \,.
\end{equation}
Equivalently speaking, $I(A,B)+I(A,E)\leq 1$ per single qubit.  This is
because Eve and Bob cannot acquire more information than is sent out
mutually by Alice and Bob whatever measurement is performed by Eve.
Therefore, in order that $I(A,B)>I(A,E)$ (Theorem 1 in
Ref.~\cite{Gisin02a}), it suffices to have $I(A,B)\geq n$.  Since
\begin{math}
I(A,B) = 2n\left[1+p_d\log_2p_d+(1-p_d)\log_2(1-p_d)\right],
\end{math}
$p_d$ is required to be less than 0.110028, approximately 11 \%, which
is the upper bound for the BB84 protocol\cite{Biham97a,Mayers93a}.  This
proves that our protocol is \emph{at least} as secure as the BB84
protocol against collective attacks.
The above lines of proof applies only for collective attacks. However,
it has been argued that the collective attack may be the optimal one of
the most general coherent attacks\cite{coherent-attacks}.
It is also interesting to note that for incoherent attacks, $I(A,E)$ in
Eq.~(\ref{qkd::eq:3}) is significantly restricted and hence the sum
$I(A,B)+I(A,E)$ per single qubit is far less than 1;
cf.~(\ref{qkd::eq:4}). This is demonstrated in Fig.~\ref{fig:4} (b)
comparing the sum for the BB84 protocol and for ours.  It suggest that
the upper bound $p_d\approx 11$ \% may be reduced further with proper
analysis of the restriction on the possible measurements by Eve.  More
detailed analysis of the security of our protocol against the coherent
attacks should therefore be an interesting topic for further studies in
the future.

\section{Impersonation Attack}
\label{sec:impersonation}

\begin{figure}
\centering
\includegraphics*[width=4cm]{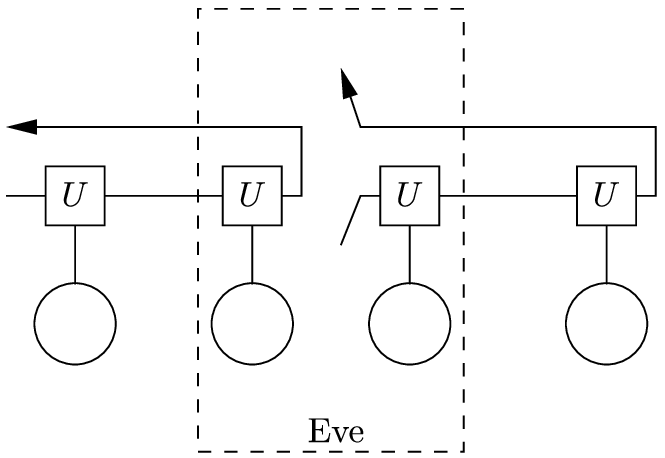}
\includegraphics*[width=4cm]{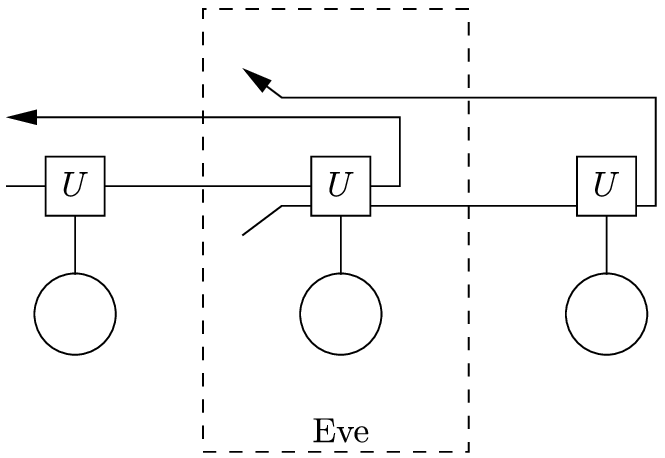}
\caption{Impersonation attack}
\label{fig:5}
\end{figure}

In our protocol, Alice sends a qubit to Bob and gets it back.  So does
Bob with another travel qubit.  It is possible for Eve to intercept the
channel and pretend to be her/his legitimate partner to each.  One can
think of two different ways of impersonation attack.  In the first
method [see Fig.~\ref{fig:5}(a)], Eve uses two home qubits of her own.
Eve can use one of the two to share a perfect key with Alice following
the procedures of the protocol in Section~\ref{sec:protocol}, and the
other to share another key with Bob.  However, the keys so generated to
Alice and Bob are independent and have no correlation.  Therefore, by
bit verification procedure, this attack can be detected with probability
$1/2$.

In the second method [see Fig.~\ref{fig:5}(b)], Eve uses only one home
qubit $E$ of her own, which is used for the interaction with both Alice and
Bob.  In this case, the total wave function of the whole qubits is given by
\begin{multline}
\label{paper::eq:10}
\left(\ket{\uparrow\uparrow\uparrow}
  +\ket{\downarrow\downarrow\downarrow}\right)
\ket{\bar\alpha\bar\beta\bar\epsilon}
+ \left(\ket{\uparrow\downarrow\uparrow}
  + \ket{\downarrow\uparrow\downarrow}\right)
\ket{\bar\alpha\beta\bar\epsilon} \\\mbox{}
+ \left(\ket{\uparrow\uparrow\downarrow}
  + \ket{\downarrow\downarrow\uparrow}\right)
\ket{\alpha\beta\epsilon}
+ \left(\ket{\uparrow\downarrow\downarrow}
  +\ket{\downarrow\uparrow\uparrow}\right)
\ket{\alpha\bar\beta\epsilon}
\end{multline}
(not normalized), where the product states are arranged such as
\begin{math}
\ket{...}_{ABE}\ket{...}_{CDE'}
\end{math}
($E'$ is the travel qubit of Eve's).  It then follows immediately that
the detection probability of this attack is still $1/2$.

\section{Photon Number Splitting Attack}
\label{sec:PNS}

\begin{figure}
\centering
\includegraphics*{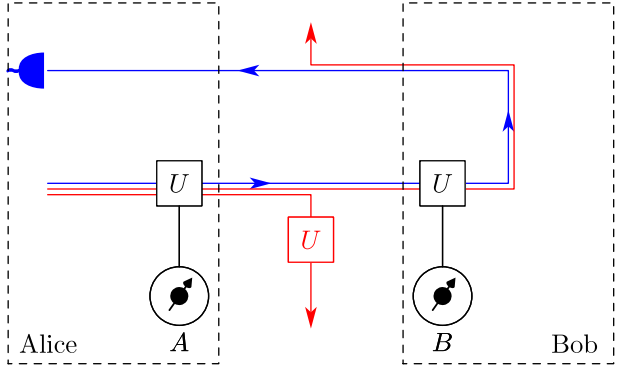}
\caption{Photon-number splitting attack.}
\label{fig:6}
\end{figure}

Finally, we investigate the security of the protocol against the photon
number splitting attack (PNS). (We note that the security analysis in
the case of lossy channel is essentially the same as that against the
PNS attack.)  Let us suppose that the photon source generate three
photons (the discussion can be trivially generalized to the case of more
photons; see below).  Eve takes one photon (say $E_1$) on the quantum
channel from Alice to Bob and another ($E_2$) on the channel back to
Alice from Bob; see Fig.~\ref{fig:6}.  Only photon $C$ finally arrives
at Alice's hand.  Similarly, Eve takes photons $E_1'$ and $E_2'$ out of
the photons from Bob.  Bob receives back only $D$.  The final state of
the whole photons and home qubits are given by
\begin{multline}
\label{PNS::eq:1}
i\ket{\uparrow\uparrow}\ket{\bar\alpha\bar\beta}
\ket{\bar\alpha\bar\beta}\ket{\bar\alpha\bar\beta}
-i\ket{\downarrow\downarrow}\ket{\alpha\beta}
\ket{\bar\alpha\bar\beta}\ket{\bar\alpha\bar\beta}
\\\mbox{}
+ \ket{\uparrow\downarrow}\ket{\bar\alpha\beta}
\ket{\alpha\beta}\ket{\alpha\beta}
+ \ket{\downarrow\uparrow}\ket{\alpha\bar\beta}
\ket{\alpha\beta}\ket{\alpha\beta} \,,
\end{multline}
where the product states have been denoted according to the arrangement
of the qubits such as
\begin{math}
\ket{..}_{AB}\ket{..}_{E_1E_2}\ket{..}_{E_1'E_2'}\ket{..}_{CD}
\end{math}.
Eve waits until Alice and Bob performs projective measurement on their
travel qubits $C$ and $D$.  Then the wave function in
Eq.~(\ref{PNS::eq:1}) collapses into either
\begin{equation}
\label{paper::eq:7}
\ket{\uparrow\uparrow}\ket{\bar\alpha\bar\beta}
\ket{\bar\alpha\bar\beta}\ket{\bar\alpha\bar\beta}
-\ket{\downarrow\downarrow}\ket{\alpha\beta}
\ket{\bar\alpha\bar\beta}\ket{\bar\alpha\bar\beta}
\end{equation}
or
\begin{equation}
\label{paper::eq:8}
\ket{\uparrow\downarrow}\ket{\bar\alpha\beta}
\ket{\alpha\beta}\ket{\alpha\beta}
+ \ket{\downarrow\uparrow}\ket{\alpha\bar\beta}
\ket{\alpha\beta}\ket{\alpha\beta} \,.
\end{equation}
Therefore, Eve can know the key without being detected simply by
checking whether
\begin{math}
\braket{\epsilon_1|\epsilon_2}\braket{\epsilon_1'|\epsilon_2'} > 0
\end{math} (Alice and Bob share the key $0$)
\begin{math}
\braket{\epsilon_1|\epsilon_2}\braket{\epsilon_1'|\epsilon_2'} < 0
\end{math} (Alice and Bob share the key $1$), where
$\epsilon_1,\epsilon_1'=\alpha,\bar\alpha$ and 
$\epsilon_2,\epsilon_2'=\beta,\bar\beta$.
This test can be easily done, for example, using an interferometer.
The discuss is trivially generalized to the case of even more photons.
It is enough for Eve to steal two photons from Alice and another two
from Bob.

\begin{figure}
\centering
\includegraphics{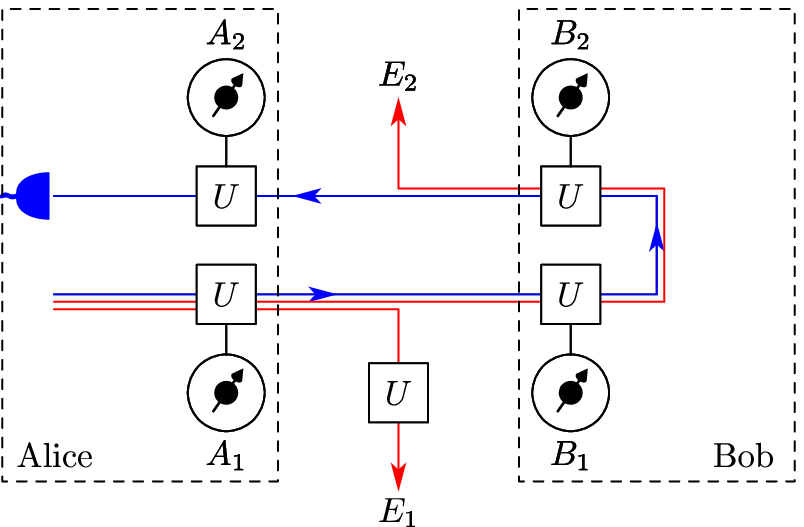}
\caption{(color on-line) A variation of the protocol using four home qubits.}
\label{fig:7}
\end{figure}

One may be tempted to overcome this problem using four home qubits (two
for Alice and two for Bob) as illustrated in Fig.~\ref{fig:7}.  This
scheme ``hides'' by means of entanglement the output state of $C$ and
$D$ even after Alice and Bob performs projective measurements on $C$ and
$D$.
However, following the similar lines as above,
the total wave function of the whole qubits is given by
\begin{widetext}
\begin{multline}
\label{paper::eq:9}
-\frac{1}{4}\ket{\bar\alpha\bar\beta}
\left(\ket{\Phi^+\Psi^+}\ket{\Phi^+\Psi^+}
  +\ket{\Phi^+\Psi^-}\ket{\Phi^+\Psi^-}
  -\ket{\Phi^-\Psi^+}\ket{\Phi^-\Psi^+}
  -\ket{\Phi^-\Psi^-}\ket{\Phi^-\Psi^-}\right)
\\\mbox{}
- \frac{1}{4}\ket{\bar\alpha\bar\beta}
\left(\ket{\Psi^-\Phi^-}\ket{\Psi^+\Phi^+}
  -\ket{\Psi^-\Phi^+}\ket{\Psi^+\Phi^-}
  +\ket{\Psi^+\Phi^-}\ket{\Psi^-\Phi^+}
  -\ket{\Psi^+\Phi^+}\ket{\Psi^-\Phi^-}\right)
\\\mbox{}
-\frac{i}{4}\ket{\alpha\beta}
\left(\ket{\Psi^-\Psi^+}\ket{\Phi^+\Phi^+}
  -\ket{\Psi^-\Psi^-}\ket{\Phi^+\Phi^-}
  -\ket{\Psi^+\Psi^+}\ket{\Phi^-\Phi^+}
  +\ket{\Psi^+\Psi^-}\ket{\Phi^-\Phi^-}\right)
\\\mbox{}
+ \frac{i}{4}\ket{\alpha\beta}
\left(\ket{\Phi^+\Phi^-}\ket{\Psi^+\Psi^+}
  +\ket{\Phi^+\Phi^+}\ket{\Psi^+\Psi^-}
  +\ket{\Phi^-\Phi^-}\ket{\Psi^-\Psi^+}
  +\ket{\Phi^-\Phi^+}\ket{\Psi^-\Psi^-}\right)
\end{multline}
\end{widetext}
arranging the product states such as
\begin{math}
\ket{..}_{CD}\ket{..}_{A_1A_2B_1B_2}\ket{..}_{E_1E_2E_1'E_2'}
\end{math}
Therefore, in order to know the key, all Eve has to do is to distinguish
the Bell state $\ket{\Phi^\pm}$ from $\ket{\Psi^\pm}$, which is as easy
as the test for the two-home-qubit scheme analyzed above.

\section{Experimental Feasibility}
\label{sec:experiment}

The parametric Faraday rotation of photon polarization by atomic spins
have been widely used in quantum optics and atomic physics.  For
example, it has been used for quantum non-demolition measurement of the
atomic spin\cite{Kitagawa93a,Wineland92a,Kuzmich00a}.  However, because
of the weak atom-photon interaction, the Faraday rotation angle is
usually quite small (several degrees).  To enhance the atom-photon
interaction to achieve the rotation angle of $\pi/2$, one has to put an
atom to a cavity.  However, trapping a single atom in a cavity is still
technologically challenging.

Another candidate for a conditional Faraday rotation of photon
polarization is the quantum dot in a micro-cavity, which has already been
demonstrated experimentally\cite{Imamoglu99a,Leuenberger}.  Here the
photon interacts with the electron spin in the semiconductor quantum
dot.  The transmission distance is limited mainly by the coherence time
of the electron spin in the quantum dot.  The maximum transmission
distance (given by the speed of light) would be ~10 m and $1 \times
10^6$ m for coherence times of 100 ns \cite{Kik} and for 10 ms
\cite{Krout} in one-way transmission.  We believe the distance limitation
will be extensively relaxed in the near future.

\section{Conclusion}
\label{sec:conclusion}

We have proposed a new QKD protocol exploring the quantum states of the
Faraday rotators.  The protocol is secure against eavesdropping for
ideal single-photon source and robust against impersonation attacks.
This protocol is not allowed for multiphoton source which produces more
than two photons.  The protocol could be implemented experimentally with
semiconductor quantum dots in micro-cavity.

\section*{Acknowledgments}
We thank J.~W. Lee and B.-G. Englert for helpful discussions.
This work was supported by the SRC/ERC program of MOST/KOSEF
(R11-2000-071), the Korea Research Foundation Grants
(KRF-2005-070-C00055 and KRF-2006-312-C00543), the SK Fund, and the
KIAS.


\begin{thebibliography}{99}
\bibitem{endnote:1} %
The initial preparation of $\ket{\phi=0}$ of QFR is just for
convenience. The initial state of QFR is required to be an arbitrary
state on the equator of the Poincar\'e sphere.
\bibitem{endnote:2} %
Yet a classical channel is necessary at the key verification state in
order to detect possible eavesdropping.
\end{thebibliography}
\end{document}